\documentclass{IEEEcsmag}

\usepackage[colorlinks,urlcolor=blue,linkcolor=blue,citecolor=blue]{hyperref}

\usepackage{upmath}

\jvol{XX}
\jnum{XX}
\paper{8}
\jmonth{May/June}
\jname{IT Professional}
\pubyear{2019}

\usepackage{hyperref}
\usepackage{url}
\usepackage{times}  
\usepackage{helvet} 
\usepackage{courier}  
\usepackage{graphicx} 
\usepackage[ruled,vlined,linesnumbered]{algorithm2e}
\usepackage{subfigure}
\newcommand{\eqbf}[1]{\boldsymbol{#1}}
\usepackage{booktabs} 
\usepackage{caption}
\usepackage{epstopdf}
\usepackage{diagbox}

\usepackage[T1]{fontenc}    
\usepackage{booktabs}       
\usepackage{amsfonts}       
\usepackage{nicefrac}       
\usepackage{amsmath}
\usepackage{graphicx}
\newcommand{\myref}[1]{(\ref{#1})}
\usepackage[ruled,vlined,linesnumbered]{algorithm2e}

\usepackage{lipsum}

\begin{document}

\sptitle{Department: Head}
\editor{Editor: Name, xxxx@email}

\title{Collaborative Generative Hashing for Marketing and Fast Cold-start Recommendation}

\author{Yan Zhang}
\affil{University of Electronic Science and Technology of China \\
	Center for Artificial Intelligence, University of Technology Sydney}

\author{Ivor W. Tsang}
\affil{Center for Artificial Intelligence, University of Technology Sydney}
\author{{L}ixin Duan*}

\affil{University of Electronic Science and Technology of China}

\markboth{Department Head}{Collaborative Generative Hashing for Marketing and Fast Cold-start Recommendation}

\begin{abstract}
Cold-start has being a critical issue in recommender systems with the explosion of data in e-commerce. Most existing studies proposed to alleviate the cold-start problem are also known as hybrid recommender systems that learn representations of users and items by combining user-item interactive and user/item content information. However, previous hybrid methods regularly suffered poor efficiency bottlenecking in online recommendations with large-scale items, because they were designed to project users and items into continuous latent space where the online recommendation is expensive. To this end, we propose a collaborative generated hashing (CGH) framework to improve the efficiency by denoting users and items as binary codes, then fast hashing search techniques can be used to speed up the online recommendation. In addition, the proposed CGH can generate potential users or items for marketing application where the generative network is designed with the principle of Minimum Description Length (MDL), which is used to learn compact and informative binary codes. Extensive experiments on two public datasets show the advantages for recommendations in various settings over competing baselines and analyze its feasibility in marketing application.	
\end{abstract}

\maketitle
\newcommand\blfootnote[1]{%
	\begingroup 
	\renewcommand\thefootnote{}\footnote{#1}%
	\addtocounter{footnote}{-1}%
	\endgroup 
}
\blfootnote{*Corresponding Author}\chapterinitial{\uppercase{New user or new item recommendation}} With the development of e-commerce, we are accustomed to receiving a variety of things we might be seeking, such as movies, books, news, hotels, etc. In the digital `world' where they are also referred to as items to be (not) recommended. Recommender systems help users find their desirable items, creating new revenue opportunities for vendors, such as Amazon, Taobao, eBay, etc. Traditional recommender systems promote items based on their content similarities, which is also known as content filtering. The drawback of content filtering is its lack of personalized information, which requires to extract  features from content information that can represent personalized information to conduct recommendation. Thus, the performance of content filtering depends on the feature learning technique. Another state-of-the-art recommendation framework is collaborative filtering where matrix factorization (MF) has been applied successfully in various scenarios. Collaborative filtering produces recommendations from user--item past interactive information. Most current recommender systems are designed according to MF. However, new items and users make the matrix factorization invalid due to the lack of interactive data, which brings a big challenge to recommender systems. Bottlenecks caused by new items or users are known as cold-start issues in recommender systems.

To solve cold-start issues, many scholars put forward hybrid recommender systems by incorporating ratings with side information, such as image, context, profile, social networks, etc. These hybrid recommendation frameworks first learn user/item representations; and then pass them into the rating prediction model (e.g. the inner product of user/item representations) to predict rating scores; and finally recommend top-$k$ items by ranking these scores. However, most existing hybrid recommender systems formulated in continuous latent space, which leads to low efficiency in the stage of online recommendation when the data size increases. Specifically, suppose the size of the item set is $m$, the time complexity of recommending top-$k$ items for a specific user in continuous space is $\mathcal{O}(mk+mlog{}k)$~\cite{zhang2017discrete}, which is a critical efficiency bottleneck with the growth of items, such as those in Taobao. It's challenging to meet the fast response requirement for a specific user.

Hashing has being a promising approach to solving the efficiency issue. To learn hash codes of users and items, existing hashing based recommendations have to optimize a nonlinear discrete problem which has proven to be NP-hard~\cite{haastad2001some}. So they transformed the original discrete optimization problem into tractable approximate surrogate problems. There are two types of approximate methods: two-stage hashing and learning-based hashing. Two-stage hashing consists of a continuous latent representation learning by discarding discrete constraints and a simple quantization method like a threshold function, which causes great information loss. Learning-based hashing obtains binary codes by optimizing approximate surrogate mixed-integer problems directly~\cite{zhang2017discrete}, which often achieves better solution for the original discrete problem, but the training of the mixed-integer problems is also expensive. 

To solve the above issues, we create hash functions by a collaborative generative network which can speed up the training procedure and reduce the quantization loss. Thus, in this paper we propose a collaborative generated hashing (CGH) to learn hash functions of users and items from content data with the principle of Minimum Description Length (MDL) that has been successfully applied to generate effective hash codes \cite{dai2017stochastic}. 

For new users and new items, manufacturers often do some marketing analysis before expanding or  producing a product. A solid marketing strategy is essential to the success of manufacturers. From the perspective of recommendation, the marketing process can be regarded as a cold-start item recommendation. Mining potential users corresponds to recommending the new item to some specific users. Thus, in our work we propose an approach to improve marketing by discovering potential users via the generative step. To reconstruct effective users, uncorrelated and balanced limits are imposed to learn compact and informative binary codes with the principle of the MDL. Specifically, for a new item, we generate a new potential user with the generative step, and then we search the nearest potential users in the user set. By recommending a new product to potential users who might be interested in but didn’t plan to buy, which is expected to attract those potential users, and thus it's a promising marketing strategy.

We organize the paper as follows: we first introduce some work closely related to CGH, and follow by introducing the framework of CGH and comparing it with related competing baselines: CDL \cite{wang2015collaborative} and DropoutNet \cite{volkovs2017dropoutnet}; we then formulate the generative step and illustrate its application in mining potential customers, leading on to the introduction of the inference step, i.e., building hash functions; we finally summarize the training objective and the optimization method, followed by the experimental analysis and the conclusion.


The main contributions of this paper are summarized as follows:
\begin{itemize}
	\item[(1)] We propose a Collaborative Generated Hashing (CGH) with the principle of MDL to learn compact but informative hash codes. This applies to various settings for recommendation.
	\item[(2)] We provide a marketing strategy by discovering potential users by the generative step of CGH. 
	\item[(3)] We evaluate the effectiveness of the proposed CGH compared with the state-of-the-art baselines, and demonstrate its robustness and convergence properties on public datasets.
\end{itemize}

\section{\uppercase{related work}}
Existing studies for solving cold-start problems were mostly modeled as a combination of collaborative filtering and content filtering, known as hybrid recommender systems \cite{wang2011collaborative,wang2015collaborative,he2016vbpr,volkovs2017dropoutnet,wang2018attention,wang2019sequential}. Specifically, they learned continuous latent factors by incorporating side information into the interactive data -- such as Collaborative Deep Learning (CDL) \cite{wang2015collaborative}, Visual Bayesian Personalized Ranking (VBPR) \cite{he2016vbpr}, Collaborative Topic modeling for Recommedation (CTR) \cite{wang2011collaborative} and the DropoutNet for addressing cold start issues\cite{volkovs2017dropoutnet}. 

Hashing techniques applied for speeding up the online recommendation fall into two groups: two-stage hashing and learning-based hashing. Two classic two-stage hashing methods are preference preserving hashing (PPH)~\cite{zhang2014preference} and iterative quantization (ITQ)~\cite{zhou2012learning}. Due to substantial information loss caused by the quantization stage, learning-based hashing frameworks were proposed to learn hash codes directly, and are composed of two types: bit-wise learning, such as discrete collaborative filtering (DCF) \cite{zhang2016discrete} and block-wise learning \cite{zhang2018discrete, wang2019adversarial}. In addition, studies of hashing method also appears in other fields, such as cycle-consistent deep generative hashing for cross-modal retrieval \cite{wu2018cycle}, generative reconstructive hashing for Video Analysis \cite{zhang2019generative}, deep semantic hashing with generative adversarial networks \cite{qiu2017deep}, etc. These generative hashing methods also motivate us to propose a generative hashing model for recommendation.

\begin{figure*}[!h]
	\centering
	\includegraphics[width=0.9\textwidth]{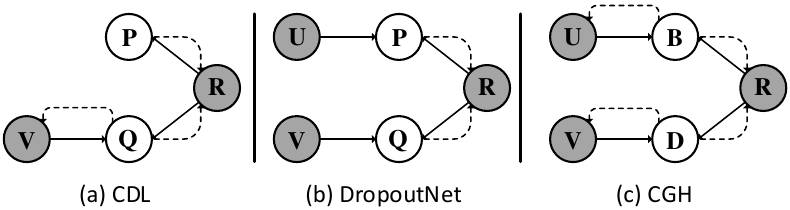}
	\caption{Differences between CDL, DropoutNet and our proposed CGH. Solid lines and dashed lines respectively represent the inference (encoding) and the generative (decoding) process. The shaded nodes $\eqbf{U}, \eqbf{V}, \eqbf{R}$ are observed user content, item content and rating, respectively. $\eqbf{P}$($\eqbf{B}$), $\eqbf{Q}$($\eqbf{D}$) denotes continuous latent factors(binary codes) of users and items.}\label{fig1}
\end{figure*}
\section{\uppercase{Collaborative generated hashing}}\label{sec2}
The framework of the proposed CGH is shown in Fig. \ref{fig1}(c), where $\eqbf{U}$, $\eqbf{V}$ and $\eqbf{R}$ are respectively observed user content, item content and rating matrix. $\eqbf{B}$ and $\eqbf{D}$ are binary codes (hash codes) of users and items, respectively. CGH consists of the generative step marked as dashed lines and the inference step denoted by solid lines. Once training is finished, we fix the model and make forward passes to obtain binary codes $\eqbf{B}$ and $\eqbf{D}$ through the inference step, and then conduct recommendation. For the marketing application, we generate a new user via the generative step for an item to search potential users who would be interested in the item.

To distinguish our proposed framework from existing associated methods, we have drawn a diagramme shown in Fig. \ref{fig1} to clarify our contributions by comparing of the collaborative deep learning (CDL) \cite{wang2015collaborative} and the DropoutNet \cite{volkovs2017dropoutnet}. In contrast to CDL, the proposed CGH aims to learn binary codes instead of continuous latent vectors $\eqbf{P}$ and $\eqbf{Q}$ because of the advantage of hashing for online recommendation; thus, the CGH optimizes an objective with the principle of MDL, while CDL optimizes the joint objective of rating loss and item content reconstruction error. In comparison with DropoutNet, CGH can be applied into marketing by discovering potential users; besides, CGH learns hash codes by the stacked denoising auto-encodoer, while DropoutNet obtains continuous latent factors from the standard neural network.

Stacked denoising auto-encoders can reconstruct data from an input of corrupted data. After giving corrupted data to the auto-encoder. It forces the hidden layer to learn robust representations; then the output will be a refined version of the input data~\cite{vincent2008extracting}. Stacked denoising auto-encoders solve this problem by corrupting the data on purpose by randomly turning some of the input values to zero. In general, the percentage of input nodes which are being set to zero depends on the amount of data and input nodes we have.


We first formulate the inference step -- constructing hash functions; we then formulate the generative process and demonstrate the marketing application -- mining potential customers; finally, we summarize the training objective and the optimization method. 


\subsection{Hash Functions}\label{sec2-2}
The inference process (also known as encoding process) shown in the Fig. \ref{fig1} (c) with solid lines, the binary latent variables $\eqbf{b}_{i}$ ($\eqbf{d}_{j}$) depends on the content vector $\eqbf{u}_{i}$ ($\eqbf{v}_{j}$) and the rating $\eqbf{R}$. Inspired by the recent work on generative hashing \cite{dai2017stochastic} and DropoutNet \cite{volkovs2017dropoutnet}, we use a multivariate Bernoulli distribution to model the inference process of $\eqbf{b}_{i}$ and $\eqbf{d}_{j}$ with linear parametrization, i.e.,
\begin{equation} \label{inf1}
	\begin{aligned}
		&q({{\eqbf{b}}_{i}}|{\tilde{\eqbf{u}}_{i}}) =\mathcal{B}(\sigma ({{\mathcal{T}}_{u}}^{T}{\tilde{\eqbf{u}}_{i}})) \\
		&q({{\eqbf{d}}_{j}}|{\tilde{\eqbf{v}}_{j}}) = \mathcal{B}(\sigma ({{\mathcal{T}}_{v}}^{T}{\tilde{\eqbf{v}}_{j}})),
	\end{aligned}
\end{equation}
where $\tilde{\eqbf{u}}_{i} = [\eqbf{u}_{i}, \eqbf{p}_{i}]$, $\tilde{\eqbf{v}}_{j} = [\eqbf{v}_{j}, \eqbf{q}_{j}]$, $\sigma(x) = (1+e^{-x})^{-1}$. $\eqbf{p}_{i}$ and $\eqbf{q}_{j}$ are the results of $r$-dimension matrix factorization \cite{koren2009matrix} of $\eqbf{R}$, i.e., $r_{ij} \approx \eqbf{p}_{i}^{T} \eqbf{q}_{j}$. $\mathcal{T}_{u} = [\eqbf{t}_{uk}]_{k=1}^{r}, \eqbf{t}_{uk} \in \mathbb{R}^{d_{u} + r}$, $\mathcal{T}_{v} = [\eqbf{t}_{vk}]_{k=1}^{r}, \eqbf{t}_{vk} \in \mathbb{R}^{d_{v} + r}$ are the output of the hidden layer in the stacked auto-encoder applied in this paper. From SGH \cite{dai2017stochastic}, the MAP solution of the eq. \myref{inf1} is readily given by
\begin{equation}\label{hash}
	\begin{aligned}
		&\eqbf{b}_{i} = \underset{\eqbf{b}_{i}}{\text{argmax}}{ \ }{q(\eqbf{b}_{i}|\eqbf{u}_{i})} = \frac{\text{sign}(\mathcal{T}_{u}^{T}\eqbf{u}_{i})+1}{2}, \\
		&\eqbf{d}_{j} = \underset{\eqbf{d}_{j}}{\text{argmax}}{ \ }{q(\eqbf{d}_{j}|\eqbf{v}_{j})} = \frac{\text{sign}(\mathcal{T}_{v}^{T}\eqbf{v}_{j})+1}{2}.
	\end{aligned}
\end{equation} 
With the linear projection followed by a sign function, we can easily get hash codes of users and items. On this basis, we also call the inference step the construction of hash functions.

\subsection{Mining Potential Customers} \label{sec2-1}
Given a sparse rating matrix $\eqbf{R}$ and item content data $\eqbf{V} \in \mathbb{R}^{d_{v}}$, where $d_{v}$ is the dimension of content vectors (bag-of-words vectors), and $\eqbf{V}$ is stacked by the bag-of-words vectors of item content in the item set $V$. Most previous studies focus on modeling deterministic frameworks to learn  representations of items, such as CDL, CTR, DropoutNet, etc. In this paper, we discover a new strategy from a perspective of marketing for item recommendation -- mining potential users.

\begin{figure*}[!h]
	\vspace{-8pt}
	\centering
	\includegraphics[width=1\textwidth]{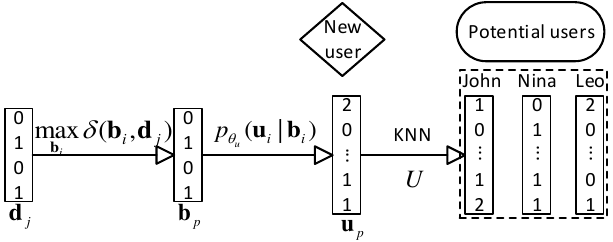}
	\caption{Demonstration of mining potential users for an item $j$. After the inference step, $\eqbf{d}_{j}$ is available, we first find out the most similar binary code $\eqbf{b}_{p}$; we then generate a new potential user $\eqbf{u}_{p}$ by the generative process; we furtherly search the top-$k$ nearest potential users from the user set by some nearest neighborhood algorithms (e.g., KNN).}\label{fig2}
	\vspace{-8pt}
\end{figure*}

We demonstrate the process of mining potential users by an item through the generative step in Fig. \ref{fig2}. Inference applies knowledge from a trained neural network model, and we use it to infer a result. In this paper, after training we obtain the trained stacked denoising auto-encoder, and then we infer items' binary representations from item content information, and we called the process of inferring as inference step or encoding process. After the inference step, the binary code of item $j$ is available. By maximizing the similarity function $\delta(\eqbf{b}_{i}, \eqbf{d}_{j})$ (also known as preference predicted model in collaborative filtering), the optimal binary code $\eqbf{b}_{p}$ is obtained. 

Given a latent representation,  we can reconstruct the corresponding input by the decoding process(also called 'generative step'). Thus, we generate a new user $\eqbf{u}_{p}$ from the binary code $\eqbf{b}_{p}$ via the generative step. Finally we discover potential users from the user set  with the aid of this new user by some nearest neighborhood algorithms such as KNN. As a marketing strategy, the generative step can detect potential users for both warm-start and cold-start items. From the perspective of marketing, it can be regarded as another kind of item recommendation.

The generative process is denoted by dashed lines in Fig. \ref{fig1} (c). Fix binary codes $\eqbf{b}_{i}$ and $\eqbf{d}_{j}$, the bag-of-words vectors $\eqbf{u}_{i}$ and $\eqbf{v}_{j}$ are generated via the generative network $p(\eqbf{\theta}_{u})$ and $p(\eqbf{\theta}_{v})$, respectively. The ratings $r_{ij}$ is generated by the similarity function of $\eqbf{b}_{i}$ and $\eqbf{d}_{j}$. We use normal Gaussian distribution to model the generation of $\eqbf{u}_{i}$ and $\eqbf{v}_{j}$ given $\eqbf{b}_{i}$ and $\eqbf{d}_{j}$, respectively:

\begin{equation}\label{ge1}
\begin{aligned}
&p({\eqbf{u}}_{i}|\eqbf{b}_{i}) =  \mathcal{N}(\mathcal{C}_{u}\eqbf{b}_{i},{{\lambda }_{u}}^{\text{-}1}\eqbf{I}\ ), \\ 
&p({\eqbf{v}}_{j}|\eqbf{d}_{j}) =  \mathcal{N}(\mathcal{C}_{v}\eqbf{d}_{j},{{\lambda }_{v}}^{\text{-}1}\eqbf{I}\ ),
\end{aligned}
\end{equation}

where $\mathcal{C}_{uk} = [\eqbf{c}_{uk}]_{k=1}^{r}, \eqbf{c}_{uk} \in \mathbb{R}^{d_{u}}$ is the codebook \cite{dai2017stochastic} with $r$ codewords. The definition of $\mathcal{C}_{v}$ is similar to $\mathcal{C}_{u}$. $d_{u}$ is the dimension of the user content vector. The prior is modeled as the multivariate Bernoulli distribution on hash codes: $p(\eqbf{b}_{i}) \sim \mathcal{B}(\rho_{u})$, and  $p(\eqbf{d}_{j}) \sim \mathcal{B}(\rho_{v})$. So, the prior probability is as follows:
\begin{equation}
\begin{aligned}
p(\eqbf{b}_{i}) = {\mathop{\prod }}_{k=1}^{r}\rho_{uk}^{{b}_{ik}}(1-\rho_{uk})^{1-{b}_{ik}}, \\
p(\eqbf{d}_{j}) = {\mathop{\prod }}_{k=1}^{r}\rho_{vk}^{{d}_{jk}}(1-\rho_{vk})^{1-{d}_{jk}},	
\end{aligned}	
\end{equation}

We predict the rating by the similarity of binary codes $\eqbf{b}_{i}$ and $\eqbf{d}_{j}$ with $\delta ({\eqbf{b}_{i}},{\eqbf{d}_{j}})=1-\frac{1}{r} \text{\textit{Hamdis}}({\eqbf{b}_{i}},{\eqbf{d}_{j}})$, where $\text{\textit{Hamdis}}({\eqbf{b}_{i}},{\eqbf{d}_{j}})$ represents the Hamming distance between hash codes, which has been extensively applied for the preference prediction in hashing-based recommender system \cite{wang2011collaborative,lian2017discrete,zhang2018discrete}. Then, the rating is thus drawn from the normal distribution centered at the predicted rating score, 
\begin{equation}
p({{r}_{ij}}|\eqbf{b}_{i}, \eqbf{d}_{j}) \sim \mathcal{N}(\delta ({\eqbf{b}_{i}},{\eqbf{d}_{j}}),C_{ij}^{-1}),
\end{equation}
where ${{C}_{ij}}$ is the precision parameter that serves as confidence of ${{r}_{ij}}$, which is similar to that in CTR \cite{wang2011collaborative} (${{C}_{ij}}=a$ if ${{r}_{ij}}=1$ and ${{C}_{ij}}=b$ otherwise). Due to the fact that ${{r}_{ij}}=0$ means the user $i$ is either not interested in item $j$ or not aware of it. 

With the generative model constructed, the joint probability of both observed ratings, content vectors and binary codes is given by 
\vspace{-4pt}
\begin{multline}
p\left(\eqbf{R},\eqbf{U},\eqbf{V},\eqbf{B},\eqbf{D} \right)=\underset{i,j}{\mathop{\prod }}\,p({{r}_{ij}}|{{\eqbf{b}}_{i}},{{\eqbf{d}}_{j}}) \\
p({{\eqbf{u}}_{i}}|{{\eqbf{b}}_{i}})p({{\eqbf{v}}_{j}}|{{\eqbf{d}}_{j}}) p({{\eqbf{b}}_{i}})p({{\eqbf{d}}_{j}})
\end{multline}

\subsection{Model Training} \label{sec2-3}
Since our goal is to reconstruct users, items and ratings by using the least information of binary codes. Thus, we train the CGH with the MDL principle, which finds the best parameters for generating hash codes that maximally compress the training data while keeping the information carried. So, CGH aims to minimize the expected amount of information related to $q$: 
\begin{equation}
\mathcal{L}(q)={{E}_{q}}[\log p(\eqbf{R},\eqbf{U},\eqbf{V},\eqbf{B},\eqbf{D})-\log q(\eqbf{B},\eqbf{D})] 
\end{equation}

Maximizing the posterior probability is equivalent to maximizing $\mathcal{L}(q)$. By simply considering the variational distribution of $q(\eqbf{B},\eqbf{D})$, the objective becomes
\begin{multline}\label{obj1}
{\mathcal{L}_{MAP}}(\Theta,\Phi) = -\sum\limits_{i,j}{\frac{{{C}_{ij}}}{2}}{{({{r}_{ij}}-\delta ({{\eqbf{b}}_{i}},{{\eqbf{d}}_{j}}))}^{2}}\\
-\frac{{{\lambda }_{u}}}{2}\sum\limits_{i}{{{({{u}_{i}}-\mathcal{C}_{v}{{\eqbf{b}}_{i}})}^{2}}} 
-\frac{{{\lambda }_{v}}}{2}\sum\limits_{j}{{{({{\eqbf{v}}_{j}}-\mathcal{C}_{v}{{\eqbf{d}}_{j}})}^{2}}} \\
 - \text{KL}({{q}_{{{\eqbf{\phi }}_{u}}}}||{{p}_{{{\eqbf{\theta }}_{u}}}})-\text{KL}({{q}_{{{\eqbf{\phi }}_{v}}}}||{{p}_{{{\eqbf{\theta }}_{v}}}})-\mathcal{r}(\Theta, \Phi)
\end{multline}
where $\Theta=\{\eqbf{\theta}_{u}, \eqbf{\theta}_{v}\}$, $\Phi = \{\eqbf{\phi}_u, \eqbf{\phi_{v}}\}$, $\mathcal{r}(\Theta, \Phi)$ is the regularizer term of parameters $\Theta$ and $\Phi$.

We train all components jointly by back-propagation. After training, we fix all parameters and make forward passes to map content data $\tilde{\eqbf{U}}$ and $\tilde{\eqbf{V}}$ to binary codes $\eqbf{B}$ and $\eqbf{D}$, respectively. The recommendation in various settings is then performed using $\eqbf{B}$ and $\eqbf{D}$ in Hamming space by ranking the preference predicted scores (similarity function) $\delta(\eqbf{b}_{i}, \eqbf{d}_{j}) = 1- \frac{1}{r}\text{Hamdis}(\eqbf{b}_{i}, \eqbf{d}_{j})$. 

The training settings depend on the recommendation settings, i.e, the warm-start setting, the cold-start item setting, and the cold-start user setting. Specifically, $-{\mathcal{L}_{MAP}}(\Theta,\Phi)$ aims to minimize the rating loss and two content reconstruction errors with regularizers. (a.) For the warm-start recommendation, less than 1 rating for all users and items are not available. Then, the above objective is optimized trivially by setting the content weights to 0 and learning hashing function with the observed ratings $\eqbf{R}$. (b.) For the cold-start item recommendation, ratings for cold-start items are missing in the training set. In this case, the objective is optimized by setting the user content weight to 0 and learning parameters with the observed ratings $\eqbf{R}$ and item content $\eqbf{V}$. (c.) The training setting for the cold-start user recommendation is similar to that for the cold-start item recommendation.

%


\section{\uppercase{Results}} \label{sec4}
We evaluate the proposed CGH on two public datasets: CiteUlike\footnote{http://www.citeulike.org/faq/data.adp} and RecSys 2017 Challenge dataset\footnote{http://www.recsyschallenge.com/2017/} from the following three aspects: 
\vspace{-8pt}
\begin{enumerate}
	\item[(1)] \textit{Online Recommendation Efficiency.}  Once training is completed, we test the online recommendation efficiency respectively in terms of real-based hybrid frameworks and hashing-based frameworks.
	\item[(2)] \textit{Marketing Analysis.} To validate the feasibility of CGH in marketing area, we first define a metric to evaluate the accuracy of mining potential users; we then test the performance separately in the warm-start and cold-start settings. 
	\item[(3)] \textit{Recommendation Accuracy.} We test the accuracy of CGH performance in various settings: the warm-start, the item cold-start, and the user cold-start with the metric Accurcy@k \cite{yin2014temporal}.
\end{enumerate}


In the following, we first introduce the experimental settings, followed by the experimental results analysis from the above contexts.
\subsection{Experimental Settings}\label{4-1}
To evaluate the effectiveness of mining potential users and the accuracy of recommendation in different settings. (1) For the CiteUlike dataset, it contains 5,551 users, 16,980 articles, 204,986 observed user-article binary interaction pairs and articles abstract content. Similar to ~\cite{wang2011collaborative}, we extract bag-of-the-words item vectors with dimension $d_{v} = 8000$ by ranking the TF-IDF values. (2) The RecSys 2017 Challenge dataset is the only publicly available dataset that contains both user and item content data, enabling both the cold-start item and cold-start user recommendation. It contains 300M user-item interactions from  1.5M users to 1.3M items and content data collected from the career-oriented social network XING (Europern analog of LinkedIn). Like \cite{volkovs2017dropoutnet}, we evaluate all methods on binary rating data (implicit feedbacks) and item content with the dimension $d_{u} = 831$, and user content with the dimension $d_{v} = 2738$.

We randomly split the implicit feedbacks (explicit ratings can be transformed into the interval of [0, 1]) $\eqbf{R}$ into three disjoint parts: warm-start ratings $\eqbf{R}^{w}$, cold-start user ratings $\eqbf{R}^{u}$, and the cold-start item ratings $\eqbf{R}^{v}$. The warm-start ratings $\eqbf{R}^{w}$ is further split into the training dataset $\eqbf{R}^{wt}$ and the testing dataset $\eqbf{R}^{we}$. Correspondingly, the user and item content datasets are split into three disjoint parts. 

In the experiments, we apply the 5-fold cross-validation method on random splits of training data and report the experimental results as the average values. The hyper-parameters settings in CGH is as follows: we set the dimension $r=50$, the layer structure of inference step is the same with DropoutNet. We set hyper-parameters for baselines according to their papers.
%

\subsection{Evaluation Metric}\label{sec4-2}
The ultimate goal of recommendation is to find out the top-$k$ items that users may be interested in. Accuracy@$k$ was widely adopted by many previous ranking-based recommender systems~\cite{koren2008factorization,chen2009collaborative}. So, we adopt the ranking-based evaluation metric Accuracy@$k$ to evaluate the quality of the recommended item ranking list.  

%

As a new application of the recommender system, there is not yet a metric to evaluate the marketing performance. Hence, in this paper, we define an evaluation metric similar to the ranking-based metric Accuracy@$k$ used for the warm-start and cold-start recommendation in this paper.


From Fig. \ref{fig2}, we discover the $k$ nearest potential users for an item $j$. The basic idea of the metric is to test whether a user in the potential users list is really interested in the item. For each positive rating ($r_{ij}=1$) in the testing dataset $D_{test}$: (1) we randomly choose 1000 negative users (users $k$ with $r_{kj}=0$) and find $k$ potential users in the 1001 user set; (2) we check if the positive user $i$ (with positive rating $r_{ij}=1$) appears in the $k$ potential users list. If the answer is `yes' we have a `hit' and a `miss' otherwise. 

The metric denoted by Accuracy@$k$ is formulated as:

\vspace{-8pt}
\begin{equation}
\small
\operatorname{Accuracy@k}=\frac{\#hit@k}{\left|D_{test}\right|},
\end{equation}
where $\left|D_{test}\right|$ is the number of positive ratings in the test set, and $\#hit@k$ denotes the number of hits in the test set.
\subsection{Online Recommendation Efficiency}
\begin{figure}
	\centering
	\includegraphics[width=1\linewidth]{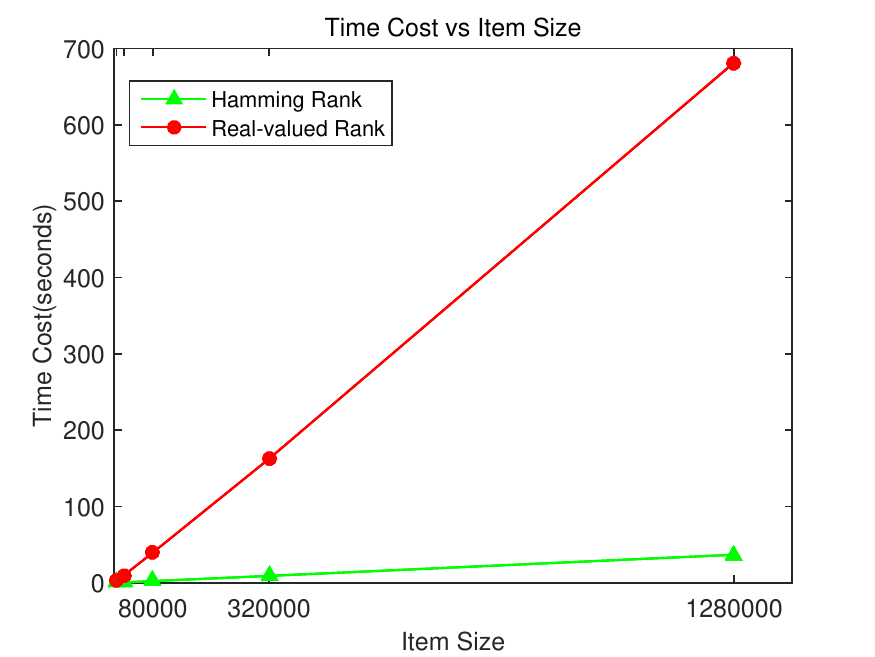}
	\caption{The time efficiency comparison of real-based recommendation frameworks and hashing-based methods.} \label{fig21}
\end{figure}

After training, continuous latent vectors and hash codes are respectively obtained by real-based hybrid recommender systems and hashing-based recommendations. We recommend top-$10$ items for a specific user separately in continuous latent space and Hamming space. We investigate the time cost on synthetic datasets for both two types of methods. We first use standard Gaussian distribution to generate items' continuous latent vectors randomly. Items hash codes are obtained from these real-valued vectors by the sign function. We set different sizes of items sets in the experiment: 80,000, 320,000, ... , 1280,000, to test the time cost of online recommendation. We denote the two kinds of online recommendations as `real-valued rank' and `Hamming rank' in Fig. \ref{fig21}. The experimental results tell us recommending in Hamming space is much more efficient than that in real space. In fact, hashing technique has been applied successfully for fast image search. It also works here because both image search and recommendation task are essentially a similarity search.
\subsection{Marketing Analysis}\label{sec4-3}
The experiments consecutively evaluate performances of marketing application by mining potential users under the warm-start and cold-start settings. Specifically, we first train the model with training dataset $\eqbf{R}^{wt}$ and the corresponding user and item content data. When the training is completed, we fix parameters and obtain hash codes $\eqbf{b}_{i}$ and $\eqbf{d}_{j}$ by making forward passes. Then we generate $k$ potential users for each item in the test dataset (the flowchart is illustrated in Fig. \ref{fig2}) and evaluate the quality of the generated potential users by the metric, `Accuracy@k'.
\begin{figure}[!htbp]
	\vspace{-4pt}
	\centering
	\includegraphics[width=1\linewidth]{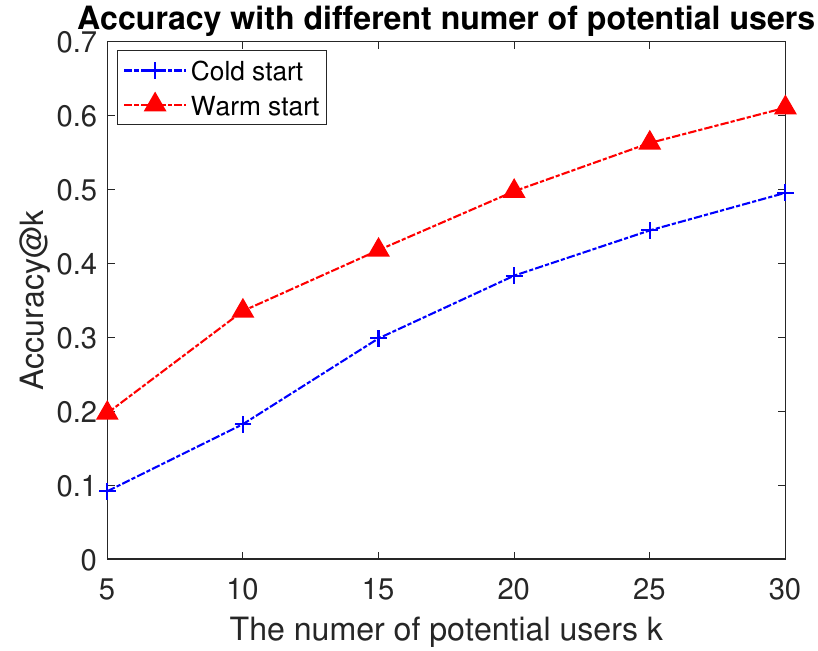}
	\caption{The marketing accuracy testing for the warm-start and cold-start settings}\label{fig31}
	\vspace{-8pt}
\end{figure}

\begin{figure}[!htbp]
	\vspace{-4pt}
	\centering
	\includegraphics[width=1\linewidth]{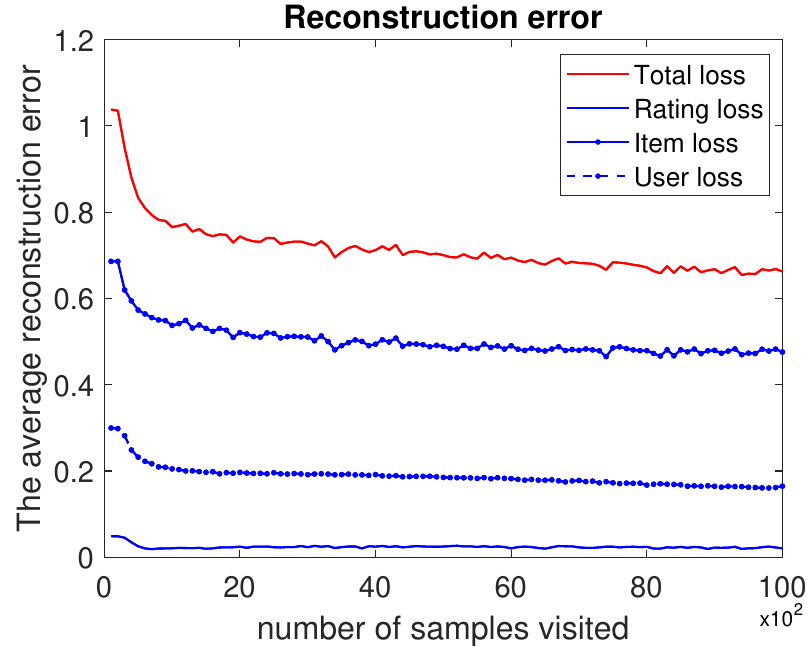}
	\caption{The average reconstruction error with the iteration training step on CiteUlike. The total error is the sum of reconstruction errors  for ratings, users' and items' content}\label{fig32}
\vspace{-8pt}
\end{figure}

The marketing analysis for warm-start items and cold-start items are reported in Fig. \ref{fig31}, which displays how the marketing accuracy varies with the number of potential users. It indicates that accuracy increases with the number of potential users for both the cold-start and warm-start settings. It's reasonable because mining more potential users increases the probability to cover the user who are really interested in the item. The accuracy in the cold-start setting is comparable to that in the warm-start setting, so CGH can be used as an approach for marketing. Moreover, for a new item, we can apply CGH to find some potential users.



%
Fig. \ref{fig32} demonstrates the convergence of CGH that reveals the training process is converged and correct.
\subsection{Recommendation Accuracy}
We assess the effectiveness of the proposed CGH by testing the accuracy of recommending top-$k$ items in Hamming space. Specifically, after training, we apply hash codes of users and items to conduct online recommendation by fast hashing technique in Hamming space. Please note that this procedure is different from the marketing application, because marketing process can also provide recommendation.

\begin{figure}
	\centering
	\includegraphics[width=1\linewidth]{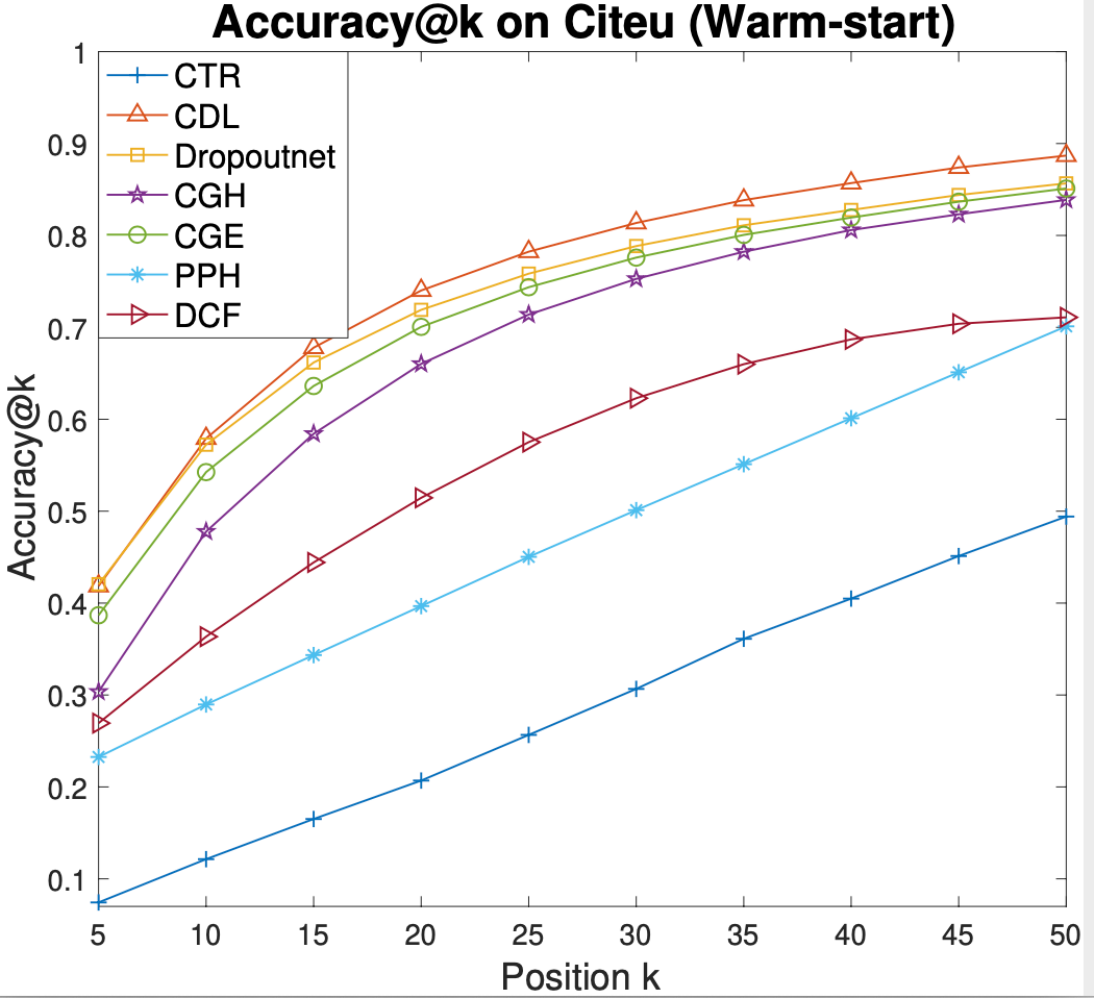}
	\caption{The accuracy comparison with the number of recommended items in the warm-start setting on CiteUlike dataset.}\label{fig41}
\end{figure}

\textbf{Accuracy for the Warm-start Recommendation.}
Fig. \ref{fig41} shows the accuracy comparison with the warm-start setting on CiteUlike dataset. In this figure, collaborative generated embedding (CGE) denotes the continuous version of the proposed CGH. The figure shows the proposed CGH (CGE) has a comparable performance with other hybrid recommender systems, and has the advantage of performance over other two hashing models PPH and DCF. It's worth noting that the proposed CGH is a hashing-based recommendation, and it has the efficiency advantage in online recommendation verified before. Due to continuous latent vectors intuitively carried more information than hash codes. So, it is acceptable to have small gaps between the real-valued hybrid recommendation and the hashing-based recommendation. 

In addition, there is still a small gap between CGE and DropoutNet, because the content reconstruction error is taken into consideration in CGH(CGE); while DropoutNet didn't consider it. In our work, the reconstruction error is significant in the generative step of CGH, which makes it a feasible approach to mining effective potential users for marketing application.

\textbf{Accuracy for the Cold-start Item Recommendation.}
We test the recommendation accuracy on the test dataset $R_{v}$ for the competing hybrid recommender systems and the proposed CGH under the same cold-start setting. Specifically, we first choose items with less than 5 ratings as cold-start items $R_{v}$, and the remaining items are regarded as warm-start items; we then train our model with warm-start ratings (i.e., ratings from warm-start users to warm-start items) and the corresponding warm-start items' content data; we finally predict ratings by the obtained users’ presentations and parameters of hash functions, and test the performance on the cold-start items $R_{v}$. Each item in $R_{v}$ owes no more than 5 positive ratings, thus these items are treated as cold-start items. Then we select users with at least one positive rating as test users. For each test user, we first choose his/her ratings related to cold-start items as the test set, and the remaining ratings as the training set. Our goal is to test whether the marked-off cold-start items can be accurately recommended to the right user. 

\begin{figure}
	\centering
	\includegraphics[width=1\linewidth]{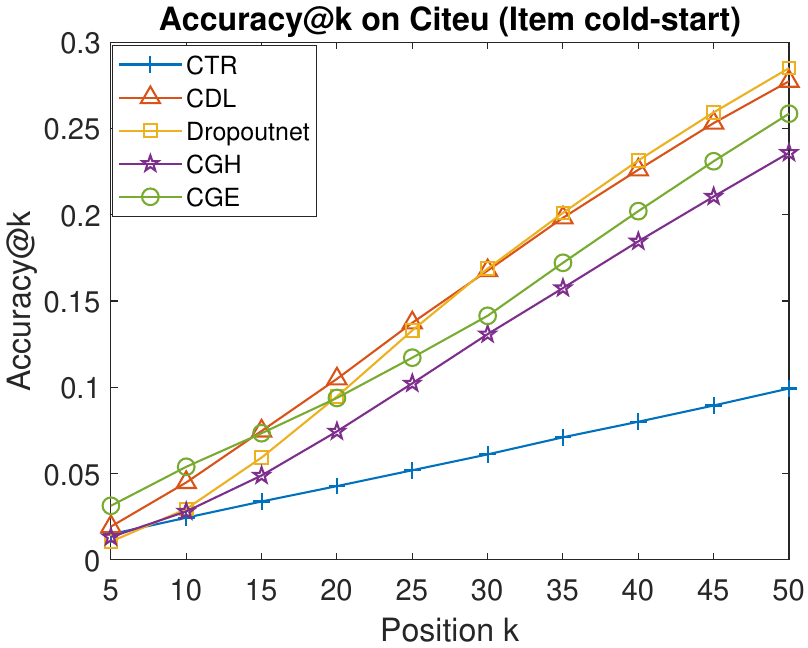}
	\caption{The accuracy variation with the number of recommended items for the warm-start setting on CiteUlike.}\label{fig42}
\end{figure}

The experimental results for the cold-start item recommendation are shown in Fig. \ref{fig42}. We conclude that CGH has a comparable performance with competing baselines and achieves better performance than CTR. The results evaluated by another metric MRR (detailed in Appendix.A) are similar.

\textbf{Accuracy for Cold-start User Recommendation.}
We also test the performance of the cold-start user setting on the test dataset $R_{u}$. Specifically, in $R_{u}$, each user (the cold-start user) has less than 5 positive ratings. Then we select items with at least one positive rating as test items. For each test item, we first choose ratings related to cold-start users as the test set, and the remaining ratings as the training set. Our goal is to test whether the test item can be can be accurately recommended to marked-off user. 
\begin{figure}
	\centering
	\includegraphics[width=1\linewidth]{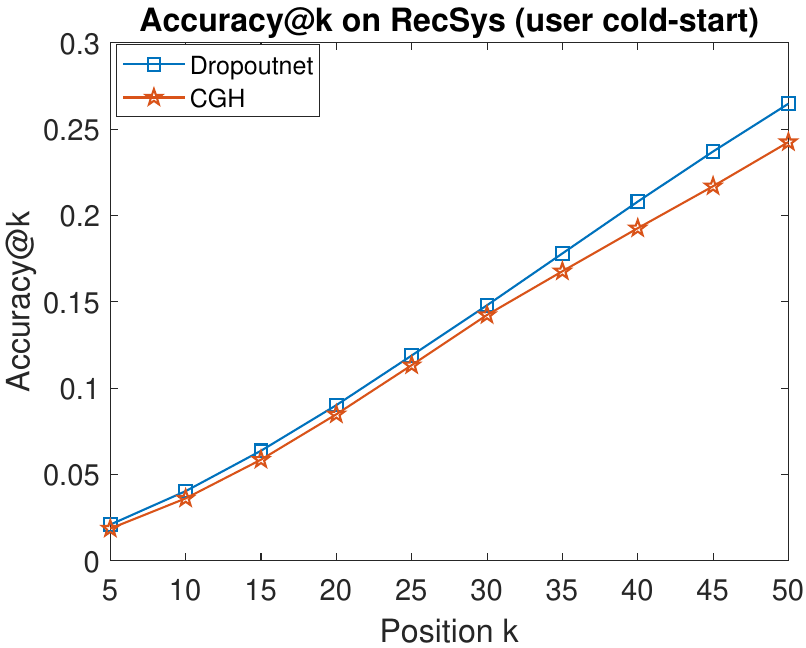}
	\caption{The accuracy variation with the number of recommended items for the warm-start setting on CiteUlike.}\label{fig43}
\end{figure}

Among baselines, only DropoutNet was designed to deal with both the cold-start user and cold-start item recommendations, so, we test the accuracy of DropoutNet and the proposed CGH under the cold-start user setting. The experimental results shown in Fig. \ref{fig43} indicate our proposed CGH can achieve comparable performance with DropoutNet. Additionally, CGH has efficient advantages in online recommendation and can be applied in marketing area.

\section{Conclusion}\label{sec5}
In this paper, a generative recommendation framework called collaborative generated hashing (CGH) is proposed to address the cold-start and efficiency issues for recommendation. The two main contributions put forward here are: (1) we develop a collaborative generated hashing framework with the principle of Minimum Description Length together (MDL) to derive compact and informative hash codes which are essential for the accurate recommendation and marketing; (2) the proposed CGH has meaningful application in marketing area by the generative step. To be precise, we design a framework to discover the $k$ potential customer using the generate step; (3) we evaluate the proposed scheme on two public datasets: the experimental results show we can achieve efficient and accurate online recommendation with hashing under both the warm-start and cold-start settings.
\section{Acknowledgment}
This work is partially supported by the Major Project for New Generation of AI under Grant No. 2018AAA0100400, the ARC under Grant DP180100106 and DP200101328. Besides, this paper is also supported by the fourth postdoctoral Innovative Talent Support Program (Grant No. BX20190061) and the 67-th China Postdoctoral Science Foundation (Grant No. 2020M673183).

\bibliography{aaai2}

\begin{thebibliography}{10}
\providecommand{\url}[1]{#1}
\csname url@samestyle\endcsname
\providecommand{\newblock}{\relax}
\providecommand{\bibinfo}[2]{#2}
\providecommand{\BIBentrySTDinterwordspacing}{\spaceskip=0pt\relax}
\providecommand{\BIBentryALTinterwordstretchfactor}{4}
\providecommand{\BIBentryALTinterwordspacing}{\spaceskip=\fontdimen2\font plus
\BIBentryALTinterwordstretchfactor\fontdimen3\font minus
  \fontdimen4\font\relax}
\providecommand{\BIBforeignlanguage}[2]{{%
\expandafter\ifx\csname l@#1\endcsname\relax
\typeout{** WARNING: IEEEtran.bst: No hyphenation pattern has been}%
\typeout{** loaded for the language `#1'. Using the pattern for}%
\typeout{** the default language instead.}%
\else
\language=\csname l@#1\endcsname
\fi
#2}}
\providecommand{\BIBdecl}{\relax}
\BIBdecl

\bibitem{zhang2017discrete}
Y.~Zhang, D.~Lian, and G.~Yang, ``Discrete personalized ranking for fast
  collaborative filtering from implicit feedback.'' in \emph{AAAI}, 2017, pp.
  1669--1675.

\bibitem{haastad2001some}
J.~H{\aa}stad, ``Some optimal inapproximability results,'' \emph{Journal of the
  ACM (JACM)}, vol.~48, no.~4, pp. 798--859, 2001.

\bibitem{dai2017stochastic}
B.~Dai, R.~Guo, S.~Kumar, N.~He, and L.~Song, ``Stochastic generative
  hashing,'' in \emph{Proceedings of the 34th International Conference on
  Machine Learning-Volume 70}.\hskip 1em plus 0.5em minus 0.4em\relax JMLR.
  org, 2017, pp. 913--922.

\bibitem{wang2015collaborative}
H.~Wang, N.~Wang, and D.-Y. Yeung, ``Collaborative deep learning for
  recommender systems,'' in \emph{Proceedings of the 21th ACM SIGKDD
  International Conference on Knowledge Discovery and Data Mining}.\hskip 1em
  plus 0.5em minus 0.4em\relax ACM, 2015, pp. 1235--1244.

\bibitem{volkovs2017dropoutnet}
M.~Volkovs, G.~Yu, and T.~Poutanen, ``Dropoutnet: Addressing cold start in
  recommender systems,'' in \emph{Advances in Neural Information Processing
  Systems}, 2017, pp. 4957--4966.

\bibitem{wang2011collaborative}
C.~Wang and D.~M. Blei, ``Collaborative topic modeling for recommending
  scientific articles,'' in \emph{Proceedings of the 17th ACM SIGKDD
  international conference on Knowledge discovery and data mining}.\hskip 1em
  plus 0.5em minus 0.4em\relax ACM, 2011, pp. 448--456.

\bibitem{he2016vbpr}
R.~He and J.~McAuley, ``Vbpr: visual bayesian personalized ranking from
  implicit feedback,'' in \emph{Thirtieth AAAI Conference on Artificial
  Intelligence}, 2016.

\bibitem{wang2018attention}
S.~Wang, L.~Hu, L.~Cao, X.~Huang, D.~Lian, and W.~Liu, ``Attention-based
  transactional context embedding for next-item recommendation,'' in \emph{32nd
  AAAI Conference on Artificial Intelligence, AAAI 2018}.\hskip 1em plus 0.5em
  minus 0.4em\relax Association for the Advancement of Artificial Intelligence,
  2018, pp. 2532--2539.

\bibitem{wang2019sequential}
S.~Wang, L.~Hu, Y.~Wang, L.~Cao, Q.~Z. Sheng, and M.~Orgun, ``Sequential
  recommender systems: challenges, progress and prospects,'' in
  \emph{Proceedings of the 28th International Joint Conference on Artificial
  Intelligence}.\hskip 1em plus 0.5em minus 0.4em\relax AAAI Press, 2019, pp.
  6332--6338.

\bibitem{zhang2014preference}
Z.~Zhang, Q.~Wang, L.~Ruan, and L.~Si, ``Preference preserving hashing for
  efficient recommendation,'' in \emph{Proceedings of SIGIR'14}.\hskip 1em plus
  0.5em minus 0.4em\relax ACM, 2014, pp. 183--192.

\bibitem{zhou2012learning}
K.~Zhou and H.~Zha, ``Learning binary codes for collaborative filtering,'' in
  \emph{Proceedings of KDD'12}.\hskip 1em plus 0.5em minus 0.4em\relax ACM,
  2012, pp. 498--506.

\bibitem{zhang2016discrete}
H.~Zhang, F.~Shen, W.~Liu, X.~He, H.~Luan, and T.-S. Chua, ``Discrete
  collaborative filtering,'' in \emph{Proceedings of SIGIR'16}, vol.~16, 2016.

\bibitem{zhang2018discrete}
Y.~Zhang, H.~Wang, D.~Lian, I.~W. Tsang, H.~Yin, and G.~Yang, ``Discrete
  ranking-based matrix factorization with self-paced learning,'' in
  \emph{Proceedings of the 24th ACM SIGKDD International Conference on
  Knowledge Discovery \& Data Mining}.\hskip 1em plus 0.5em minus 0.4em\relax
  ACM, 2018, pp. 2758--2767.

\bibitem{wang2019adversarial}
H.~Wang, N.~Shao, and D.~Lian, ``Adversarial binary collaborative filtering for
  implicit feedback,'' in \emph{Proceedings of the AAAI Conference on
  Artificial Intelligence}, vol.~33, 2019, pp. 5248--5255.

\bibitem{wu2018cycle}
L.~Wu, Y.~Wang, and L.~Shao, ``Cycle-consistent deep generative hashing for
  cross-modal retrieval,'' \emph{IEEE Transactions on Image Processing},
  vol.~28, no.~4, pp. 1602--1612, 2018.

\bibitem{zhang2019generative}
J.~Zhang, Z.~Wei, I.~C. Duta, F.~Shen, L.~Liu, F.~Zhu, X.~Xu, L.~Shao, and
  H.~T. Shen, ``Generative reconstructive hashing for incomplete video
  analysis,'' in \emph{Proceedings of the 27th ACM International Conference on
  Multimedia}, 2019, pp. 845--854.

\bibitem{qiu2017deep}
Z.~Qiu, Y.~Pan, T.~Yao, and T.~Mei, ``Deep semantic hashing with generative
  adversarial networks,'' in \emph{Proceedings of the 40th International ACM
  SIGIR Conference on Research and Development in Information Retrieval}, 2017,
  pp. 225--234.

\bibitem{vincent2008extracting}
P.~Vincent, H.~Larochelle, Y.~Bengio, and P.-A. Manzagol, ``Extracting and
  composing robust features with denoising autoencoders,'' in \emph{Proceedings
  of the 25th international conference on Machine learning}.\hskip 1em plus
  0.5em minus 0.4em\relax ACM, 2008, pp. 1096--1103.

\bibitem{koren2009matrix}
Y.~Koren, R.~Bell, and C.~Volinsky, ``Matrix factorization techniques for
  recommender systems,'' \emph{Computer}, no.~8, pp. 30--37, 2009.

\bibitem{lian2017discrete}
D.~Lian, R.~Liu, Y.~Ge, K.~Zheng, X.~Xie, and L.~Cao, ``Discrete content-aware
  matrix factorization,'' in \emph{Proceedings of KDD'17}.\hskip 1em plus 0.5em
  minus 0.4em\relax ACM, 2017, pp. 325--334.

\bibitem{yin2014temporal}
H.~Yin, B.~Cui, L.~Chen, Z.~Hu, and Z.~Huang, ``A temporal context-aware model
  for user behavior modeling in social media systems,'' in \emph{Proceedings of
  the 2014 ACM SIGMOD international conference on Management of data}.\hskip
  1em plus 0.5em minus 0.4em\relax ACM, 2014, pp. 1543--1554.

\bibitem{koren2008factorization}
Y.~Koren, ``Factorization meets the neighborhood: a multifaceted collaborative
  filtering model,'' in \emph{Proceedings of the 14th ACM SIGKDD international
  conference on Knowledge discovery and data mining}.\hskip 1em plus 0.5em
  minus 0.4em\relax ACM, 2008, pp. 426--434.

\bibitem{chen2009collaborative}
W.-Y. Chen, J.-C. Chu, J.~Luan, H.~Bai, Y.~Wang, and E.~Y. Chang,
  ``Collaborative filtering for orkut communities: discovery of user latent
  behavior,'' in \emph{Proceedings of the 18th international conference on
  World wide web}.\hskip 1em plus 0.5em minus 0.4em\relax ACM, 2009, pp.
  681--690.

\end{thebibliography}
\bibliographystyle{ieeetrans}

\section{\uppercase{Appendix: another metric experiments}}
We evaluate the accuracy in terms of the MRR \cite{yin2014temporal} metric shown in Table \ref{tab:3} for the warm-start recommendation. Our proposed CGH performs as well as the best result of the real-valued competing baselines. Table~\ref{tab:3} summarizes MRR results for the four algorithms: the best result is marked as \lq${\star}$\rq~ and the second best is marked as~\lq${o}$\rq~. We find that the performance of CGH is very close to the best result, which is consistent with the outcome of Accuracy@$k$ reported in Fig. \ref{fig41}. 


\begin{table}[!htbp]
	\centering
	\caption{MRR on CiteUlike}
	%
	\begin{tabular}{c|c|c|c|c}
		\hline
		\textbf{Method} & \textbf{CTR} & \textbf{CDL} & \textbf{Dropoutnet} & \textbf{CGH} \\
		\hline
		\lq Warm-start\rq &0.0324 & 0.0667$^{\star}$ & 0.0580 & 0.0595$^{o}$ \\
		\hline
		\lq Cold-start\rq &0.0101 & 0.0150 & 0.0179$^{\star}$ & 0.0165$^{o}$  \\
		\hline
	\end{tabular}
	\label{tab:3}
\end{table}

\begin{IEEEbiography}{Yan Zhang}{\,}received the B.S. degree from the Sichuan Normal University in 2012, and the Ph.D. degree from the University of Electronic Science and Technology of China in 2019. Her research interests include recommender system, machine learning, and logic synthesis.
\end{IEEEbiography}


\begin{IEEEbiography}{Ivor W. Tsang}{\,}is an ARC Future Fellow and Professor of Artificial Intelligence, at University of Technology Sydney (UTS). He is also the Research Director of the Australian Artificial Intelligence Institute (AAII). He has 200 research papers published in top-tier journals and conferences. According to Google Scholar, he has more than 15,000 citations and his H-index is 57. In 2013, Prof Tsang received his prestigious Australian Research Council Future Fellowship for his research regarding Machine Learning on Big Data. In 2019, his Journal of Machine Learning Research paper titled "Towards ultrahigh dimensional feature selection for big data" received the International Consortium of Chinese Mathematicians Best Paper Award. In 2020, Prof Tsang was recognized as the AI 2000 AAAI/IJCAI Most Influential Scholar in Australia for his outstanding contributions to the field of AAAI/IJCAI between 2009 and 2019. In addition, he had received the  IEEE Transactions on Neural Networks Outstanding 2004 Paper Award in 2007, the 2014 IEEE Transactions on Multimedia Prize Paper Award, and a number of best paper awards and honors from reputable international conferences, including the Best Student Paper Award at CVPR 2010. He serves as a Senior Area Chair/Area Chair for NeurIPS, ICML, AISTATS and IJCAI. He serves as the Editorial Board for the Journal of Machine Learning Research and Machine Learning Journal, and serves as an Associate Editor for the IEEE Transactions on Big Data and the IEEE Transactions on Emerging Topics in Computational Intelligence.
\end{IEEEbiography}


\begin{IEEEbiography}{Lixin Duan}{\,}received the B.Eng. degree from the University of Science and Technology of China in 2008, and the Ph.D. degree from Nanyang Technological University in 2012. He is currently a Full Professor with the School of Computer Science and Engineering, University of Electronic Science and Technology of China (UESTC). His research interests include machine learning algorithms (especially in transfer learning and domain adaptation) and their applications in computer vision. He was a recipient of the Microsoft Research Asia Fellowship in 2009 and the Best Student Paper Award at the IEEE Conference on Computer Vision and Pattern Recognition 2010.
\end{IEEEbiography}

\end{document}